\documentclass[11pt,a4paper]{article}
\usepackage[hyperref]{naaclhlt2019}
\usepackage{times}
\usepackage{latexsym}

\usepackage{url}
\usepackage{amsmath}
\usepackage{amssymb}
\usepackage{graphicx}
\usepackage{subfigure}
\aclfinalcopy 

\title{Graph Convolution for Multimodal Information Extraction \\ 
from Visually Rich Documents}

\author{Xiaojing Liu, Feiyu Gao, Qiong Zhang, Huasha Zhao\\
  Alibaba Group \\
  {\tt \{huqiang.lxj,feiyu.gfy,qz.zhang,huasha.zhao\}@alibaba-inc.com}}

\date{}

\begin{document}
\maketitle
\begin{abstract}
Visually rich documents (VRDs) are ubiquitous in daily business and life. Examples are purchase receipts, insurance policy documents, custom declaration forms and so on. In VRDs, visual and layout information is critical for document understanding, and texts in such documents cannot be serialized into the one-dimensional sequence without losing information. Classic information extraction models such as BiLSTM-CRF typically operate on text sequences and do not incorporate visual features. In this paper, we introduce a graph convolution based model to combine textual and visual information presented in VRDs. Graph embeddings are trained to summarize the context of a text segment in the document, and further combined with text embeddings for entity extraction. Extensive experiments have been conducted to show that our method outperforms BiLSTM-CRF baselines by significant margins, on two real-world datasets. Additionally, ablation studies are also performed to evaluate the effectiveness of each component of our model.

\end{abstract}

\section{Introduction}

Information Extraction (IE) is the process of extracting structured information from unstructured documents. IE is a classic and fundamental Natural Language Processing (NLP) task, and extensive research has been made in this area. Traditionally, IE research focuses on extracting entities and relationships from plain texts, where information is primarily expressed in the format of natural language text. However, a large amount of information remains untapped in VRDs.

VRDs present information in the form of both text and vision. The semantic structure of the document is not only determined by the text within it but also the visual features such as layout, tabular structure and font size of the document. Examples of VRDs are purchase receipts, insurance policy documents, custom declaration forms and so on. Figure 1 shows example VRDs and example entities to extract.

VRDs can be represented as a graph of {\em text segments} (Figure 2), where each text segment is comprised of the position of the segment and the text within it. The position of the text segment is determined by the four coordinates that generate the bounding box of the text. There are other potentially useful visual features in VRDs, such as fonts and colors, which are complementary to the position of the text. They are out of the scope of this paper, and we leave them to future works.

\begin{figure}[t]
    \centering
    \subfigure[Purchase receipt]{
    \begin{minipage}[b]{0.22\textwidth}
        \includegraphics[width=\textwidth]{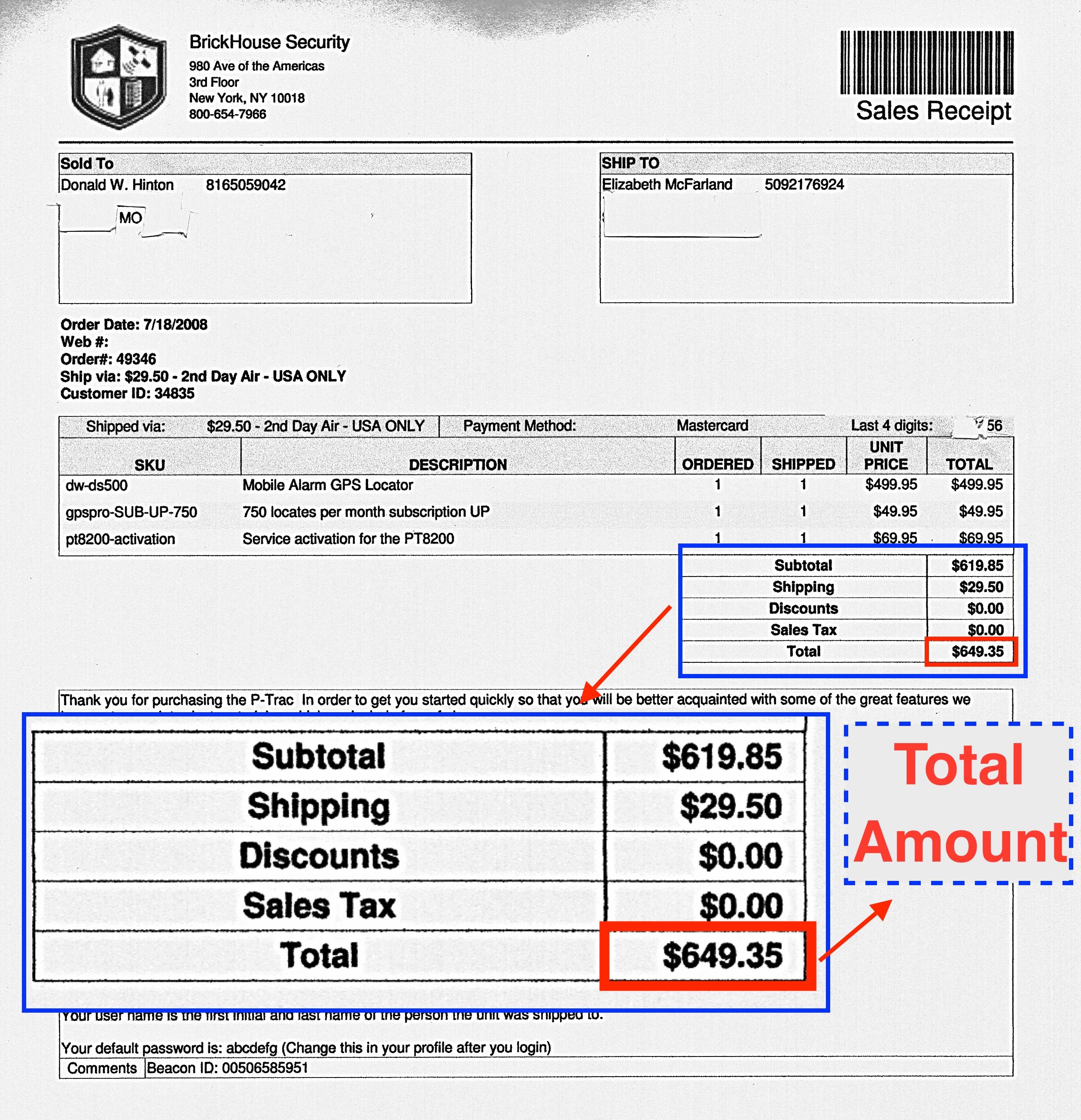}
        \label{receipt}
    \end{minipage}}
    \subfigure[Value-added tax invoice]{
    \begin{minipage}[b]{0.24\textwidth}
        \includegraphics[width=\textwidth]{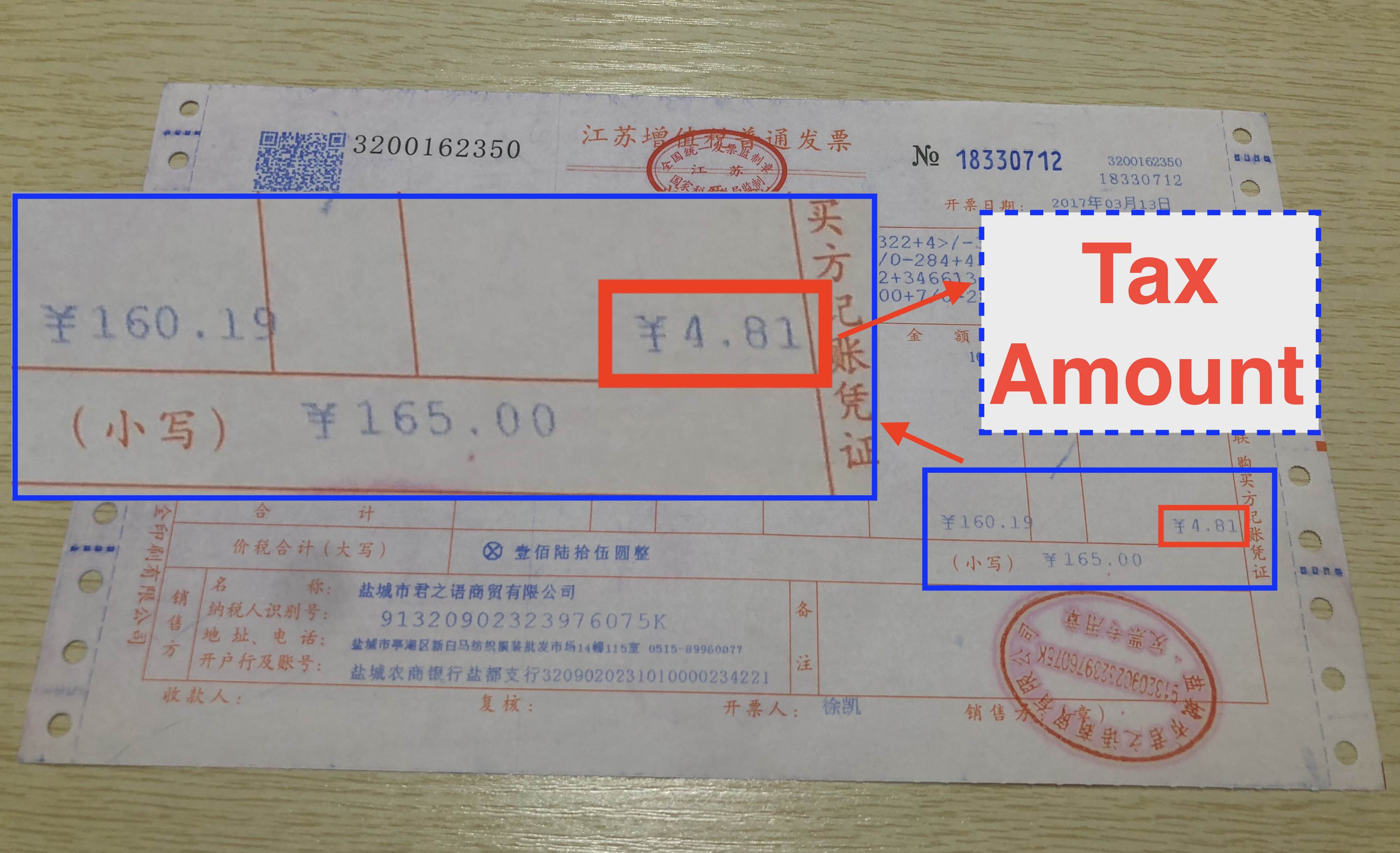}
        \label{vat}
    \end{minipage}}    
    \label{fig:examples}
    \caption{Examples of VRDs and example entities to extract.}
\end{figure}

The problem we address in this paper is to extract the values of pre-defined entities from VRDs.  We propose a graph convolution based method to combine textual and visual information presented in VRDs. The graph embeddings produced by graph convolution summarize the context of a text segment in the document, which are further combined with text embeddings for entity extraction using a standard BiLSTM-CRF model. The following paragraphs summarize the challenges of the task and the contributions of our work.

\subsection{Challenges}

IE from VRDs is a challenging task, and the difficulties mainly arise from how to effectively incorporate visual cues from the document and the scalability of the task.

First, text alone is not adequate to represent the semantic meaning in VRDs, and the contexts of the texts are usually expressed in visual cues. For example, there might be multiple dates in the purchase receipts. However, it is up to the visual part of the model to distinguish between the Invoice Date, Transaction Date, and Due Date. Another example is the tax amount in value-added tax invoice, as shown in Figure 1(b). There are multiple ``money'' entities in the document, and there is a lack of any textual context to determine which one is tax amount. To extract the tax amount correctly, we have to leverage the (relative) position of the text segment and visual features of the document in general.

Template matching based algorithms \cite{chiticariu2013rule,dengel2002smartfix,schuster2013intellix} utilize visual features of the document to extract entities; however, we argue that they are mostly not scalable for the task in real-world business settings. There are easily thousands of vendors on the market, and the templates of purchase receipts from each vendor are not the same. Thousands of templates need to be created and maintained in this single scenario. It requires substantial efforts to update the template and make sure it's not conflicting with the rest every time a new template comes in, and the process is error-prone. Besides, user uploaded pictures introduce another dimension of variance from the template. An example is shown in Figure 1(b). Value-added tax invoice is a nation-wide tax document, and the layout is fixed. However, pictures taken by users are usually distorted, often blurred and sometimes contain interfering objects in the image. A simple template-based system performs poorly in such a scenario, while sophisticated rules require significant engineering efforts for each scenario, which we believe is not scalable.

\subsection{Contributions}

In this paper, we present a novel method for IE from VRDs. The method first computes graph embeddings for each text segment in the document using graph convolution. The graph embeddings represent the context of the current text segment where the convolution operation combines both textual and visual features of the context. Then the graph embeddings are combined with text embeddings to feed into a standard BiLSTM for information extraction.

Extensive experiments have been conducted to show our method outperforms BiLSTM-CRF baselines by significant margins, on two real-world datasets. Additionally, ablation studies are also performed to evaluate the effectiveness of each component of our model. Furthermore, we also provide analysis and intuitions on why and how individual components work in our experiments.

\section{Related Works}

Our work is inspired by recent research in the area of graph convolution and information extraction.

\subsection{Graph Convolution Network} 

Neural network architectures such as CNN and RNNs have demonstrated huge success on many artificial intelligence tasks where the underlying data has grid-like or sequential structure \cite{krizhevsky2012imagenet,kim2016character,kim2014convolutional}. Recently, there is a surge of interest in studying the neural network structure operating on graphs \cite{kipf2016semi,hamilton2017inductive}, since much data in the real world is naturally represented as graphs. Many works attempt to generalize convolution on the graph structure. Some use a spectrum based approach where the learned model depends on the structure of the graph. As a result, the approach does not work well on dynamic graph structures. The others define convolution directly on the graph \cite{velivckovic2017graph,hamilton2017inductive,xu2018graph2seq,johnson2018image,duvenaud2015convolutional}. We follow the latter approach in our work to model the text segment graph of VRDs.

Different from existing works, this paper introduces explicit edge embeddings into the graph convolution network, which models the relationship between vertices directly. Similar to \cite{velivckovic2017graph}, we apply self-attention \cite{vaswani2017attention} to define convolution on variable-sized neighbors, and the approach is computationally efficient since the operation is parallelizable across node pairs. 

\subsection{Information Extraction}

Recently, significant progress has been made in information extraction from unstructured or semi-structured text. However, most works focus on plain text documents \cite{peng2017cross,lample2016neural,ma2016end,chiu2016named}. For information extraction from VRDs, \cite{palm2017cloudscan} which uses a recurrent neural network (RNN) to extract entities of interest from VRDs (invoices) is the closest to our work, but does not take visual features into account. Besides, some of the studies \cite{d2018field,medvet2011probabilistic,rusinol2013field} in the area of document understanding deal with a similar problem to our work, and explore using visual features to aid text extraction from VRDs; however, approaches they proposed are based on a large amount of heuristic knowledge and human-designed features, as well as limited in known templates, which are not scalable in real-world business settings. We also acknowledge a concurrent work of \cite{katti2018chargrid}, which models 2-D document using convolution networks. However, there are several key differences. Our neural network architecture is graph-based, and our model operates on text segments instead of characters as in \cite{katti2018chargrid}.  

Besides, information extraction based on the graph structure has been developed most recently. \cite{peng2017cross,song2018n} present a graph LSTM to capture various dependencies among the input words and \cite{wang2018joint} designs a novel graph schema to extract entities and relations jointly. However, their models are not concerned with visual information directly.

\newcommand{\norm}[1]{\left\lVert#1\right\rVert}

\section{Model Architecture}

This section describes the document model and the architecture of our proposed model. Our model first encodes each text segment in the document into graph embedding, using multiple layers of graph convolution. The embedding represents the information in the text segment given its visual and textual context. By visual context, we refer to the layout of the document and relative positions of the individual segment to other segments. Textual context is the aggregate of text information in the document overall; our model learns to assign higher weights on texts from neighbor segments. Then we combine the graph embeddings with text embeddings and apply a standard BiLSTM-CRF model for entity extraction.

\begin{figure}
    \centering
     \includegraphics[width=0.45 \textwidth]{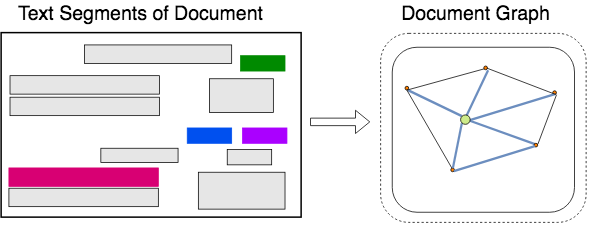}
    \label{documentgraph}
    \caption{Document graph. Every node in the graph is fully connected to each other.}
\end{figure}

\subsection{Document Modeling}
\label{documentgraph}
We model each document as a graph of text segments (see Figure 2), where each text segment is comprised of the position of the segment and the text within it. The graph is comprised of nodes that represent text segments,  and edges that represent visual dependencies, such as relative shapes and distance, between two nodes. Text segments are generated using an in-house Optical Character Recognition (OCR) system.

Mathematically, a document $\mathcal{D}$  is a tuple $(T, E)$, where $T = \{t_1, t_2, \cdots, t_n\}, t_i \in \mathcal{T}$ is a set of $n$ text boxes/nodes, $R = \{r_{i1}, r_{i2}, \cdots, r_{ij}\}, r_{ij} \in \mathcal{R}$ is a set of edges, and $E = T \times R \times T$  is a set of directed edges of the form $(t_i, r_{ij}, t_j)$ where $t_i, t_j \in T$ and $r_{ij} \in R$. In our experiments, every node is connected to each other.

\begin{figure*}[t!]
    \centering
     \includegraphics[width=0.95 \textwidth]{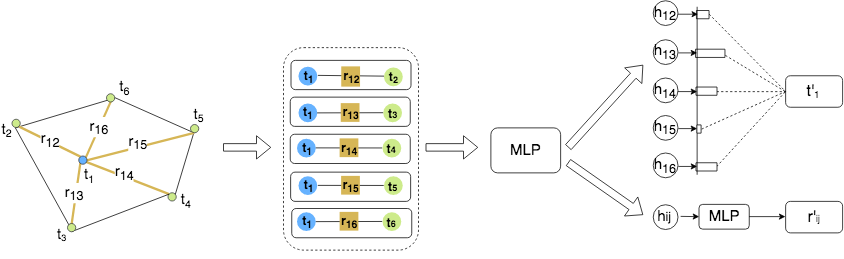}
    \label{graphconv}
    \caption{Graph convolution of document graph. Convolution is defined on  node-edge-node triplets $(t_i, r_{ij}, t_j)$. Each layer produces new embeddings for both nodes and edges.}
\end{figure*}

\subsection{Feature Extraction} 

For node $t_i$, we calculate node embedding $\mathbf{t}_i$ using a single layer Bi-LSTM \cite{schuster1997bidirectional} to extract features from the text content in the segment. 

Edge embedding between node $t_i$ and node $t_j$ is defined as follows,

\begin{align}
\mathbf{r}_{ij} = [x_{ij}, y_{ij}, \frac{w_{i}}{h_{i}}, \frac{h_{j}}{ h_{i}}, \frac{w_{j}}{h_{i}}],
\end{align}

where $x_{ij}$ and $y_{ij}$ are horizontal and vertical distance between the two text boxes respectively, and $w_i$ and $h_i$ are the width and height of the corresponding text box. The third, fourth and fifth value of the embedding are the aspect ratio of node $t_i$, relative height, and width of node $t_j$ respectively. Empirically, a visual distance between two segments is an important feature. For example, in general, the positions of relevant information are closer in one document, such as the key and value of an entity. Moreover, the shape of the text segment plays a critical role in representing semantic meanings. For example, the length of the text segment which has address information is usually longer than that of one which has a buyer name. Therefore, we use edge embedding to encode information regarding the visual distance between two segments, the shape of the source node, and the relative size of the destination node.

To summarize, node embedding encodes textual features, while edge embedding primarily represents visual features.

\subsection{Graph Convolution}
\label{graphconv}
Graph convolution is applied to compute visual text embeddings of text segments in the graph, as shown in Figure 3. Different from existing works, we define convolution on the node-edge-node triplets $(t_i, r_{ij}, t_j)$ instead of on the node alone. We compare the performances of the models using nodes only and node-edge-node triplets in Section 5.3. For node $t_i$, we extract features $\mathbf{h}_{ij}$ for each neighbour $t_j$ using a multi-layer perceptron (MLP) network, 

\begin{align}
\mathbf{h}_{ij} = g(\mathbf{t}_i,  \mathbf{r}_{ij}, \mathbf{t}_j) = \text{MLP}([\mathbf{t}_i  \Vert  \mathbf{r}_{ij}  \Vert \mathbf{t}_j]), 
\end{align}

where $ \Vert$ is the concatenate operation. There are several benefits of using this triplet feature set. First, it combines visual features directly into the neighbor representation. Furthermore, the information of the current node is copied across the neighbors. As a result, the neighbor features can potentially learn where to attend given the current node.

In our model, graph convolution is defined based on the self-attention mechanism. The idea is to compute the output hidden representation of each node by attending to its neighbors. In its most general form, each node can attend to all the other nodes, assuming a fully connected graph.

Concretely the output embedding $\mathbf{t}_i^\prime$ of the layer for node $t_i$ is computed by, 
\begin{align}
\mathbf{t}_i^\prime = \sigma(\sum_{j \in \{1, \cdots, n\}} \alpha_{ij} \mathbf{h}_{ij}),
\end{align}

where $\alpha_{ij}$ are the attention coefficients, and $\sigma$ is an activation function. In our experiments, the attention mechanism is designed as the follows, 

\begin{align}
\alpha_{ij} = \frac {\exp({\text{LeakyRelu}(\mathbf{w}_a^T  \mathbf{h}_{ij})})} {\sum_{j \in  \{1, \cdots, n\}} \exp({\text{LeakyRelu}(\mathbf{w}_a^T  \mathbf{h}_{ij})})}  ,
\end{align}

where $\mathbf{w}_a$ is a shared attention weight vector. We apply the LeakyRelu activation function to avoid the ``dying Relu'' problem and to increase the ``contrast'' of the attention coefficients potentially.

The edge embedding output of the graph convolution layer is defined as, 

\begin{align}
\mathbf{r}_{ij}^\prime = \text{MLP}(\mathbf{h}_{ij} ).
\end{align}

Outputs $\mathbf{t}_i^\prime$, $\mathbf{r}_{ij}^\prime$ are fed as inputs to the next layer of graph convolution (as computed in equation 2) or network modules for downstream tasks.

\begin{figure}
    \centering
     \includegraphics[width=0.45 \textwidth]{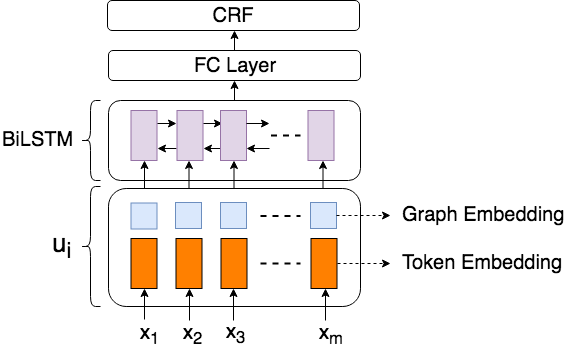}
    \label{bilstmcrf}
    \caption{BiLSTM-CRF with graph embeddings.}
\end{figure}

\subsection{BiLSTM-CRF with Graph Embeddings}
\label{bilstmcrf}

We combine graph embeddings with token embeddings and feed them into standard BiLSTM-CRF for entity extraction. As illustrated in Figure 4, visual text embedding generated from the graph convolution layers of the current text segment is concatenated to each token embedding of the input sequence; Intuitively, graph embedding adds contextual information to the input sequence. 

Formally, assume for node $t_i$, the input token sequence of the text segment is $x_1, x_2, \cdots, x_m$, and the graph embedding of the node is $\mathbf{t}_i^\prime$. The input embedding $\mathbf{u}_i$ is defined as,

\begin{align}
\mathbf{u}_i =  e(x_i) \Vert \mathbf{t}_i^\prime
\end{align}

where $e$ is token embedding lookup function, and Word2Vec vectors are used as token embeddings in our experiments.

Then the input embeddings are fed into a Bi-LSTM network to be encoded, and the output is further passed to a fully connected network and then a CRF layer.

\section{Model Supervision and Training}

We build an annotation system to facilitate the labeling of the ground truth data. For each document, we label the values for each pre-defined entity, and their locations (bounding boxes). To generate training data, we first identify the text segment each entity belongs to, and then we label the text in the segment according to IOB tagging format \cite{sang1999representing}. We assign label O to all tokens in empty text segments.

Since human annotated bounding boxes cannot match OCR detected box exactly, we apply a simple heuristic to determine which text segment an entity belongs to based on overlap area. A text segment is considered to contain an entity if $A_{overlap} / \min({A_{annotator}, A_{ocr}})$ is bigger than a manually set threshold; Here $A_{annotator}, A_{ocr}, A_{overlap}$ are the area of the annotated box of the corresponding entity, the area of the ocr detected box and the area of the overlap between the two boxes respectively. 

In our experiments, the graph convolution layers and BiLSTM-CRF extractors are trained jointly. Furthermore, to improve prediction accuracy, we add the segment classification task which classifies each text segment into a pre-defined tag as an auxiliary task and discuss the effect of multi-task learning in Section 5.4.3. We feed the graph embedding of each text segment into a sigmoid classifier to predict the tag. Since the parameters of the graph convolution layers are shared across the extraction task and segment classification task, we employ a multi-task learning approach for model training. In multi-task training, the goal is to optimize the weighted sum of the two losses. In our experiments, the weight is determined using a principled approach as described in \cite{kendall2017multi}. The idea is to adjust each task's relative weight in the loss function by considering task-dependant uncertainties.

\section{Experiments}

We apply our model for information extraction from two real-world datasets. They are Value-Added Tax Invoices (VATI) and International Purchase Receipts (IPR). 

\subsection{Datasets Description}
\label{datasets}

VATI consists of 3000 user-uploaded pictures and has 16 entities to exact. Example entities are the names of buyer/seller, date and tax amount. The invoices are in Chinese, and it has a fixed template since it is national standard invoice. However, there are many noises in the documents which include distracting objects in the image and skewed document orientation to name a few. IPR is a data set of 1500 scanned receipt documents in English which has 4 entities to exact (Invoice Number, Vendor Name, Payer Name and Total Amount). There exist 146 templates for the receipts. Variable templates introduce additional difficulties to the IPR dataset. For both datasets, we assign 70\% of each dataset for training, 15\% for validation and 15\% for the test.  The number of text segments varies per document from 100 to 300.

\subsection{Baselines}

We compare the performance of our system with two BiLSTM-CRF baselines. Baseline I applies BiLSTM-CRF to each text segment, where each text segment is an individual sentence. Baseline II applies the tagging model to the concatenated document.  Text segments in a document are concatenated from left to right and from top to bottom according to \cite{palm2017cloudscan}. Baseline II incorporates a one-dimensional textual context to the model.

\subsection{Results}
\label{res}

\begin{table}
    \centering
        \begin{tabular}{|l|c|c|c|}
        \hline
        Model         & VATI &IPR \\
        \hline
        Baseline I   & 0.745 & 0.747  \\
        Baseline II    & 0.854 & 0.820   \\
        BiLSTM-CRF + GCN  & 0.873 & 0.836  \\
         \hline
        \end{tabular}
    \caption{$F_1$ score. Performance comparisons.}
    \label{tag}
\end{table}

\begin{table}
    \centering
        \begin{tabular}{|l|c|c|c|}
        \hline
        Entities         & Baseline I & Baseline II & Our model \\
        \hline
        Invoice \#    & 0.952 & 0.961 & 0.975  \\
        Date    & 0.962 & 0.963 & 0.963  \\
        Price          & 0.527 & 0.910 & 0.943  \\
        Tax      & 0.584 & 0.902 & 0.924  \\
        Buyer   & 0.402 & 0.797 & 0.833  \\
        Seller   & 0.681 & 0.731 & 0.782  \\
        \hline
        \end{tabular}
    \caption{$F_1$ score. Performance comparisons for individual entities from VATI dataset.}
    \label{tag}
\end{table}

We use the $F_1$ score to evaluate the performances of our model in all experiments. The main results are shown in Table 1. As we can see, our proposed model outperforms both baselines by significant margins. Capturing patterns from VRDs with one-dimensional text sequence is difficult. More specifically, we present performance comparisons of six entities from VATI dataset in Table 2. It can be seen that compared with two baselines, our model performs almost identical on ``simple'' entities which can be distinguished by the text segment's text feature alone ($i.e$., Invoice Number and Date) where visual features and context information are not necessary. However, our proposed model clearly outperforms baselines on entities which can not be represented by text alone, such as Price, Tax, Buyer, and Seller. 

To further examine the contributions made by each sub-component of the graph convolution network, we perform the following ablation studies. In each study, we exclude visual features (edge embeddings), textual features (node embeddings) and the use of attention mechanism respectively, to see their impacts on $F_1$ scores on both two datasets. As presented in Table 3, it can be seen that visual features play a critical role in the performance of our model; they lead to more than 5\% performance drop in both datasets. Intuitively, visual features provide more information about contexts of the text segments, so it improves the performance by discriminating between text segments with similar semantic meanings. Moreover, textual features make similar contributions. Furthermore, the attention mechanism shows more effectiveness on variable template datasets, which results in 1.5\% performance gains. However, it makes no contribution to fixed layout datasets. We make further discussions on attention in the next section.

\begin{table}
    \centering
        \begin{tabular}{|l|c|c|}
        \hline
        Configurations/Datasets & VATI & IPR \\
        \hline
        Full model  & 0.873 &  0.836 \\
        w/o vis. features & 0.808  & 0.775 \\
        w/o text features  & 0.871 &  0.817 \\
        w/o attention           & 0.872 & 0.821 \\
      \hline
        \end{tabular}
    \caption{$F_1$ score. Ablation studies of individual component of graph convolution.}
    \label{ablation}
\end{table}

\subsection{Discussions}

\subsubsection{Attention Analysis}
To better understand how attention works in our model, we study the attention weights (on all other text segments) of each text segment in the document graph. Interestingly, for the VATI dataset, we find that attention weights are usually concentrated on a fixed text segment with strong textual features (the segment contains address information) regardless of the text segment studied. The reason behind that may be attention mechanism tries to find an anchor point of the document, and one anchor is enough as VATI documents all share the same template. Strong textual features help locate the anchor more accurately. 

For variable layout documents, more attention is paid to nearby text segments, which reflects the local structure of the document. Specifically, attention weights of the left and upper segments are even higher. Furthermore, segments that contain slot keywords such as ``address'', ``amount'' and ``name'' receive higher attention weights. 
    
\subsubsection{The Number of Graph Convolution Layers}
\label{gcnlayer}
Here we evaluate the impact of different numbers of graph convolution layers. In theory, a higher number of layers encodes more information and therefore can model more complex relationships. We perform the experiments on selected entity types of the VATI dataset, and the results are presented in Table 4. As we can see, additional layers in the network do not help simple tasks. By simple task, we mean that task achieves high accuracy with single layer graph convolution. However, more layers indeed improve the performances of more difficult tasks. As shown in the table, the optimal number of layers is two for our task, as three layers overfit the model. Ideally, the number of graph convolution layers used should be adaptive to the specific task, of which we leave the study to future works.

\begin{table}
    \centering
        \begin{tabular}{|l|c|c|c|}
        \hline
        Entities         & 1 layer & 2 layer & 3 layer \\
        \hline
        Invoice \#    & 0.959  & 0.975 & 0.964 \\
        Date    & 0.960 & 0.963 &0.960 \\
        Price          & 0.931  & 0.943 & 0.931\\
        Tax      & 0.915  & 0.924 & 0.917\\
        Buyer   & 0.829  & 0.833  & 0.827\\
        Seller   & 0.772& 0.782  & 0.775\\
        \hline
        \end{tabular}
    \caption{$F_1$ score. Performance comparisons of different graph convolution layers for individual entities from VATI dataset.}
    \label{tag}
\end{table}

\subsubsection{Multi-Task Learning}
As shown in Table 5, our task benefits from the segment classification task and multi-task learning method in both datasets. The two tasks in our experiments are complementary, and compared with the single task model, the multi-task learning model may have better generalization performance by adopting more information. Furthermore, we find that the multi-task learning helps the training converge much faster.

\begin{table}
    \centering
        \begin{tabular}{|c|c|c|c|}
        \hline
        Model         & VATI &IPR \\
        \hline
        BiLSTM-CRF + GCN  & 0.873 & 0.836  \\
          + Multi-task    & 0.881 & 0.849  \\
         \hline
        \end{tabular}
    \caption{$F_1$ score. Effectiveness of multi-task learning approach.}
    \label{tag}

\end{table}

\section{Conclusions and Future Works}
This paper studies the problem of entity extraction from VRDs. A graph convolution architecture is proposed to encode text embeddings given visually rich context. BiLSTM-CRF is applied to extract the final results. We manually annotated two real-world datasets of VRDs, and perform comprehensive experiments and analysis. Our system outperforms BiLSTM baselines and presents a novel method for IE from VRDs. Furthermore, we plan to extend the graph convolution framework to other tasks in VRDs, such as document classification.

\section*{Acknowledgments}

We thank the OCR team of Alibaba Group for providing us OCR service support and anonymous reviewers for their helpful comments.  

\bibliographystyle{acl_natbib}

\bibliography{naaclhlt2019_camera}

\end{document}